\documentclass[twocolumn,aps,pra,10pt,showpacs]{revtex4-2}
\usepackage{bm}
\usepackage{amsfonts}
\usepackage{amssymb}
\usepackage{amsmath}
\usepackage{graphicx}
\begin{document}

\title{Backflow in relativistic wave equations}
\author{Iwo Bialynicki-Birula}\email{birula@cft.edu.pl}
\affiliation{Center for Theoretical Physics, Polish Academy of Sciences\\
Aleja Lotnik\'ow 32/46, 02-668 Warsaw, Poland}
\author{Zofia Bialynicka-Birula}
\affiliation{Institute of Physics, Polish Academy of Sciences\\
Aleja Lotnik\'ow 32/46, 02-668 Warsaw, Poland}
\author{Szymon Augustynowicz}
\affiliation{Faculty of Physics, University of Warsaw, Pasteura 5, 02-093 Warsaw, Poland}
\date{\today}

\begin{abstract}
We show that, contrary to the statements made by many authors, the backflow is not a nonclassical effect. The backflow is a characteristic feature of solutions of the wave equations: quantum and classical. We present simple solutions of the Dirac equation, Maxwell equations and equations of linearized gravity where the backflow phenomenon is clearly seen. In this work we describe backflow in relativistic theories but this phenomenon can occur in the solutions of all kinds of wave equations: quantum and classical.
\end{abstract}

\maketitle

\section{Introduction}

The aim of this work is to question the following statements about the backflow that associate this phenomenon exclusively with quantum theory: ``peculiar quantum effect'' \cite{bm}, ``the backflow effect is the intriguing quantum-mechanical phenomenon'' \cite{hetal}, ``this surprising and clearly nonclassical effect'' \cite{year}, ``backflow is clearly a nonclassical effect'' \cite{yh}, ``quantum backflow is a classically impossible phenomenon'' \cite{bg}, `` new quantum effect'' \cite{b}, or ``Quantum backflow, discovered quite a few years back, is a generic purely quantum phenomenon'' \cite{bisg}. Most recent papers on the subject also refer to backflow as ``quantum backflow'' suggesting that backflow is inseparably connected with quantum theory. We dispel all these opinions by showing that the backflow effect is seen in full glory not only in the Dirac theory but also in such typically classical theories as Maxwell electrodynamics and linearized theory of gravity.  Backflow is a wave phenomenon which occurs in all kinds of wave theories: quantum and classical. The authors of the ``classically forbidden'' explanations apparently have not read the paper by Berry \cite{mb} who has shown that the backflow effect also exist in simple one-dimensional situations in classical optics. We extend the results obtained by Berry to the full-blown Maxwell electrodynamics.

In general terms the phenomenon of backflow is the counterintuitive behavior of the flow of some quantity (energy, probability, etc.). Namely, in some regions of space the direction of the flow is {\em opposite} with respect to the direction of all its constituent elementary waves. In quantum wave mechanics studied in \cite{bm,hetal,year,yh,bg,b,bisg} the quantity under study is the probability density whose flow is determined by the probability current. The constituent elementary waves are the plane waves appearing in the Fourier decomposition of the wave function. We underscore the universality of the backflow by using several examples  of beams that exhibit this phenomenon in all three cases considered here: in Dirac theory, in Maxwell theory, and in the linearized theory of gravity. Our examples are: a simple superposition of two monochromatic plane waves, exponential beams, Bessel beams, and the hopfion solutions. In all these examples we will not pay any attention to the overall value of the wave amplitude because the analysis of the backflow relies only on the relative values of the flow: positive vs. negative.

\section{Dirac theory}

We begin with the relativistic theory of electrons because this is close to the nonrelativistic wave mechanics studied in \cite{bm,hetal,year,yh,bg,b}. The simplest solution of the Dirac equation which exhibits the phenomenon of backflow is the superposition of two monochromatic plane waves. This solution is easily generated by differentiation from the solution of the Klein-Gordon equation according to the procedure explained in \cite{bb0,bb1}. The solution of the Klein-Gordon equation in the form of a plane waves subjected to this procedure gives the following solution of the Dirac equation $(c=1,\,\hbar=1)$,
\begin{align}\label{dpw}
\psi_{\bm p}(\bm r,t)=\left[\begin{array}{c}m\\0\\E_p-p_z\\p_x+ip_y\end{array}\right]e^{-iE_p t+i{\bm p}\cdot{\bm r}}.
\end{align}
The sum of two plane waves,
\begin{align}\label{dpw1}
\psi(\bm r,t)=\psi_{\bm p}(\bm r,t)+\psi_{\bm q}(\bm r,t),
\end{align}
produces the following $z$ component of the current,
\begin{align}\label{d2pw}
j_z(\bm r,t)&=\psi^*(\bm r,t)\alpha_z(\psi(\bm r,t)\nonumber\\
&=\frac{2}{m^2}\Big(p_z(E_p-p_z)+q_z(E_q-q_z)\nonumber\\
&+(m^2+E_pq_z+E_qp_z-E_pE_q)\cos\big(\phi(\bm r,t)\big)\nonumber\\
&+(p_yq_x-p_xq_y)\sin\big(\phi(\bm r,t)\big)\Big),
\end{align}
where $\alpha_z$ is the Dirac matrix in the Weyl representation of $\gamma$ matrices \cite{weyl} and $\phi(\bm r,t)=(E_p-E_q)t-(\bm{p-q})\cdot{\bm r}$. The plot of $j_z$ presented in Fig.1 shows the presence of backflow in the solution (\ref{dpw1}) of the Dirac equation.
\begin{figure}
\begin{center}
\includegraphics[width=0.47\textwidth,
height=0.3\textheight]{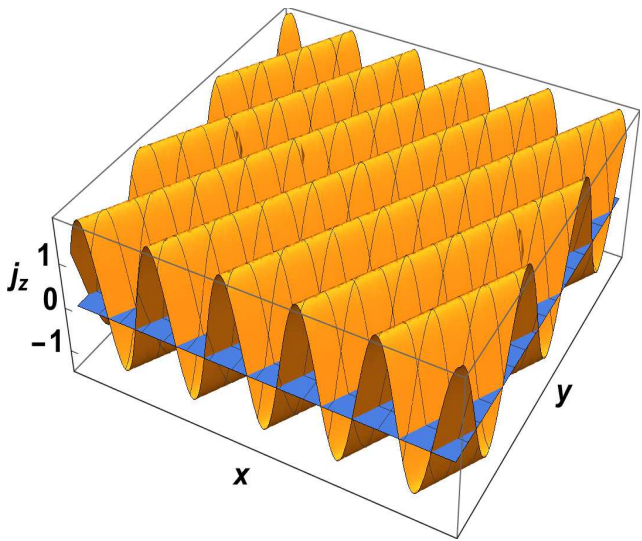}
\caption{The $z$ component of the current for the solution (\ref{d2pw}) of the Dirac equation. The $j_z=0$ plane clearly marks the regions of negative values of $j_z$. This plot is obtained for the following values of the two momenta: $p_x=0.5,\,p_y=-0.5,\,p_z=0.1,\,q_x=-0.5,\,q_y=0.5,\,q_z=0.1$.}\label{fig1}
\end{center}
\end{figure}

\section{Maxwell theory}

In our study of backflow in Maxwell theory we use the Riemann-Silberstein (RS) vector $\bm F(\bm r,t)$ which significantly simplifies the calculations \cite{rs},
\begin{align}\label{rs}
\bm F(\bm r,t)=\!\frac{{\bm D}(\bm r,t)}{\sqrt{2\epsilon_0}}+i\frac{{\bm B}(\bm r,t)}{\sqrt{2\mu_0}}.
\end{align}
In terms of $\bm F(\bm r,t)$ Maxwell equations take on a concise form ($c=1$),
\begin{align}\label{max0}
i\partial_t{\bm F}(\bm r,t)={\bm\nabla}\times{\bm F}(\bm r,t),\quad{\bm\nabla}\cdot{\bm F}(\bm r,t)=0.
\end{align}
Separation of these equations into their real and imaginary parts, leads to the standard Maxwell equations.

As the simplest example of the electromagnetic field, which will be shown to exhibit backflow, we choose again the superposition of two monochromatic plane waves. The plane-wave solution of Maxwell equations may be written in the form,
\begin{align}\label{pw}
{\bm F}_{\bm k}(\bm r,t)=
{\bm e}(\phi_{\bm k},\theta_{\bm k})e^{i(\bm{k}\cdot\bm{r}-kt)}.
\end{align}
where $k=|\bm{k}|$ and the polarization vector is
\begin{align}\label{ee}
{\bm e}(\phi,\theta)=
\frac{1}{\sqrt{2}}\left[\begin{array}{c}\vspace{0.2cm}
e^{2i\phi}\sin^2(\theta/2)-\cos^2(\theta/2)\\\vspace{0.2cm}
-ie^{2i\phi}\sin^2(\theta/2)-i\cos^2(\theta/2)\\
2e^{i\phi}\sin(\theta/2)\cos(\theta/2)\end{array}
\right].
\end{align}
This polarization vector is the same as the one defined by Eq.(22) of \cite{str}, expressed in spherical coordinates,
\begin{align}\label{sph}
k_x=k\cos\phi\sin\theta,\,k_y=k\sin\phi\sin\theta,\,
k_z=k\cos\theta.
\end{align}
\begin{figure}
\begin{center}
\includegraphics[width=0.47\textwidth,
height=0.3\textheight]{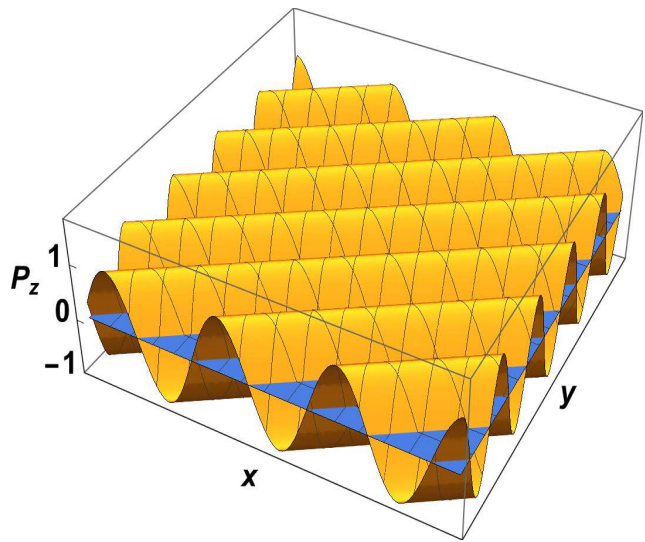}
\caption{The $z$ component of the Poynting vector for the solution (\ref{rs2}) of the Maxwell equations. This plot was obtained with the following values of the parameters: $k=1,\,l=1,\,\phi_k=\pi/2,\,\phi_l=0,\,\theta_k=\pi/2,\,
\theta_l=\pi/2$. The $P_z=0$ plane clearly marks the regions where $P_z$ has negative values.}
\label{fig2}
\end{center}
\end{figure}
The RS vector describing the superposition of two plane waves with wave vectors $\bm k$ and $\bm l$ is simply the sum,
\begin{align}\label{rs2}
{\bm F}(\bm r,t)={\bm e}(\phi_{\bm k},\theta_{\bm k})e^{i(\bm{k}\cdot\bm{r}-kt)}+
{\bm e}(\phi_{\bm l},\theta_{\bm l})e^{i(\bm{l}\cdot\bm{r}-lt)}.
\end{align}
In order to exhibit the phenomenon of backflow, we will assume that the $z$ components of both wave vectors $\bm k$ and $\bm l$ are positive, i.e. $0<\theta<\pi/2$.

In quantum wave mechanics the backflow is associated with the counterintuitive behavior of the probability current. In electromagnetism, the role of the quantum probability density is played by the energy density $\mathcal E$ and the role the quantum probability current is played by the energy current, described by the Poynting vector ${\bm P}={\bm E}\times{\bm H}$.

In the context of the backflow the correspondence between wave mechanics and electromagnetism is very close. In both cases, the density is positive and the continuity equation is satisfied,
\begin{align}\label{cont}
\partial_t\mathcal{E}+\bm\nabla\!\cdot\!\bm{P}=0.
\end{align}

The formulas for the energy density and the Poynting vector written in terms of the RS vector are \cite{rs},
\begin{align}
\qquad\mathcal{E}(\bm r,t)&={\bm F}^*(\bm r,t)\!\cdot\!{\bm F}(\bm r,t),\\
\qquad\bm{P}(\bm r,t)&=-i({\bm F}^*(\bm r,t)\times{\bm F}(\bm r,t)).\label{genb}
\end{align}

To address the phenomenon of backflow we need only the $z$ component of the Poynting vector. Choosing the RS vector (\ref{rs2}), we obtain,
\begin{align}\label{poyz}
P_z&=C_k-S_k+C_l-S_l\nonumber\\
&+2C_kC_l\cos\Delta_1
-2S_kS_l\cos(\Delta_1-2\Delta_2),
\end{align}
where
\begin{subequations}
\begin{align}\label{not}
C_k&=\cos^2(\theta_k/2),\;S_k=\sin^2(\theta_k/2),\nonumber\\
C_l&=\cos^2(\theta_l/2),\;S_l=\sin^2(\theta_l/2),\\
\Delta_1&=(\bm{k}\cdot\bm{r}-kt)-(\bm{l}\cdot\bm{r}-lt),\;
\Delta_2=\phi_k-\phi_l.
\end{align}
\end{subequations}
The form of the expression for $P_z$ indicates that the best chance for its negative value is when both terms in the second line of (\ref{poyz}) are negative. This means that $\Delta_2=\pi/2$. Since $\cos(\Delta_1)$ varies between 1 and -1, the condition for the backflow is,
\begin{align}\label{cond}
R(\theta_k,\theta_l)=\frac{2C_kC_l+2S_kS_l}
{C_k-S_k+C_l-S_l}>1.
\end{align}
This condition is satisfied for all values of $\theta_k$ and $\theta_l$ in the range $\{0,\pi/2\}$. The choice of parameters  specified in the figure caption leads to a very simple result,
\begin{align}\label{pzm}
P_z=\cos\left(k(x-y)\right),
\end{align}
depicted in Fig.~2. The Poynting vector, like the probability current in quantum theory \cite{bm}-\cite{b} and in the Dirac theory, in some regions of space has negative $z$ component even though both plane waves have positive components of $p_z$. This very simple example shows that the backflow is a purely wave phenomenon not necessarily connected with quantum theory. A more elaborate example in the form of a localized wave-packet of the electromagnetic field, which exhibits backflow, is described in Appendix A.

There are two important questions concerning the electromagnetic backflow. What is the velocity of the backflow? Can backflow propagate with the speed of light? In order to find answers to these questions, we must first define the velocity of the electromagnetic radiation. Since the energy flux is determined by the Poynting vector, the ratio of this vector to the energy density defines the local velocity of propagation,
\begin{align}\label{vel}
{\bm v}(\bm r,t)=\frac{\bm{P}(\bm r,t)}{\mathcal{E}(\bm r,t)}
\end{align}
The square of the velocity has a simple form when expressed in terms of the RS vector,
\begin{align}\label{vel1}
\frac{{\bm v}^2}{c^2}=1-\frac{|\bm{F}\!\cdot\!\bm{F}|^2}
{(\bm{F}^*\!\cdot\!\bm{F})^2}.
\end{align}
Therefore, the velocity attains the speed of light when $\bm{F}\!\cdot\!\bm{F}=0$, i.e. when both field invariants vanish. In our case of two plane waves, this condition becomes,
\begin{align}\label{cond1}
{\bm e}(\phi_{\bm k},\theta_{\bm k})\!\cdot\!{\bm e}(\phi_{\bm l},\theta_{\bm l})=0,
\end{align}
and it gives,
\begin{align}\label{cond2}
\cos(\phi_{\bm k}-\phi_{\bm l})\sin(\theta_{\bm k})\sin(\theta_{\bm l})+\cos(\theta_{\bm k})\cos(\theta_{\bm l})=1.
\end{align}
This condition can be easily satisfied. This does not mean that the wave can propagate with the speed of light because in then the $x$ and $y$ components of the Poynting vector vanish. In this case the backflow disappears since we have just one plane wave propagating in the $z$ direction. However, $v_z$ can become arbitrarily close to the speed of light.

\section{Linearized gravity}

A simple way to describe the backflow in linearized gravity is by making use of the analogy between electromagnetism and gravity \cite{mb1}. The counterpart of the RS vector is the symmetric complex tensor $\mathcal{G}_{ij}$ built from the components of the Riemann tensor \cite{bb},
\begin{align}\label{sdg}
\mathcal{G}_{ij}=R_{0i0j}
+\frac{i}{2}\epsilon_{0ikl}R_{0j}^{\;\;\;kl}.
\end{align}
This tensor is directly related to the Bel-Robinson tensor \cite{mb1,ibb}. The counterpart of the electromagnetic field energy is the super-energy,
\begin{align}\label{sen}
\mathcal{E}^G(\bm r,t)=\sum_{ij}\mathcal{G}_{ij}^*(\bm r,t)\mathcal{G}_{ij}(\bm r,t),
\end{align}
and the counterpart of the Poynting vector is the super-Poynting vector \cite{mb,ibb}
\begin{align}\label{pzg}
P^G_i(\bm r,t)=\frac{i}{2}\sum_{jkl}\varepsilon_{ijk}\mathcal{G}_{jl}^*(\bm r,t)\mathcal{G}_{kl}(\bm r,t),
\end{align}
\begin{figure}
\begin{center}
\includegraphics[width=0.47\textwidth,
height=0.3\textheight]{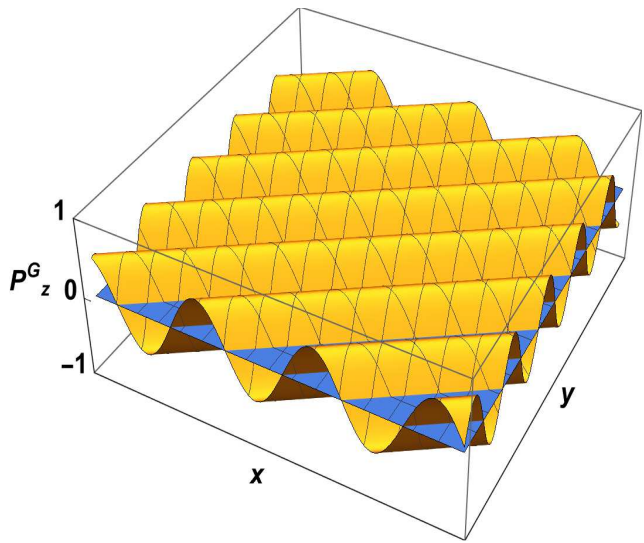}
\caption{The $z$ component of the super-Poynting vector for the solution (\ref{gpw}) of the equations of linearized gravity. This plot is obtained for the same values of the parameters as in Fig.2. The $P^G_z=0$ plane clearly marks the regions of negative values of the $z$ component of the super Poynting vector in linearized gravity. }\label{fig3}
\end{center}
\end{figure}

The tensor $\mathcal{G}$ describing the superposition of two plane waves is \cite{bb}
\begin{align}\label{gpw}
\mathcal{G}_{ij}(\bm r,t)=e_i(\phi_{\bm k},\theta_{\bm k})e_j(\phi_{\bm k},\theta_{\bm k})e^{i(\bm{k}\cdot\bm{r}-kt)}\nonumber\\
+e_i(\phi_{\bm l},\theta_{\bm l})e_j(\phi_{\bm l},\theta_{\bm l})e^{i(\bm{l}\cdot\bm{r}-lt)},
\end{align}
where ${\bm e}(\phi_{\bm k}$ is the same polarization vector as in the electromagnetic case. The substitution of this expression into (\ref{pzg}) in the general case gives a very complicated formula. However, by choosing the wave vectors as those in the electromagnetic case, we obtain again a very simple formula for the $z$ component of the super-Poynting vector,
\begin{align}\label{gpz}
P^G_z=-\frac{1}{2}\sin\left(k(x-y)\right),
\end{align}
plotted in Fig.3.

The analysis of the backflow velocity that was carried out for the electromagnetic field applies also to the gravitational waves in the linearized theory. The backflow can propagate with the velocity arbitrarily close to the speed of light. In the case of the solutions of the Dirac equation the backflow velocity $j_z/\rho$ may approach the speed of light only in ultra-relativistic case, when the mass can be neglected in comparison with momenta.

\section{Beams with fixed value of momentum in the $z$ direction}

In this section we describe the beams that are exceptionally well suited for the analysis of the backflow: the exponential beams and Bessel beams. Exponential beams were introduced in \cite{bb2} for the solutions of Maxwell equations and in \cite{bb0} for the Dirac wave functions. Bessel beams are often used in optics to describe beams carrying angular momentum. Since the exponential and Bessel beams have a well defined component of the wave vector (momentum) in the $z$-direction, the presence of backflow can be easily spotted.
\begin{figure}
\begin{center}
\includegraphics[width=0.47\textwidth,
height=0.22\textheight]{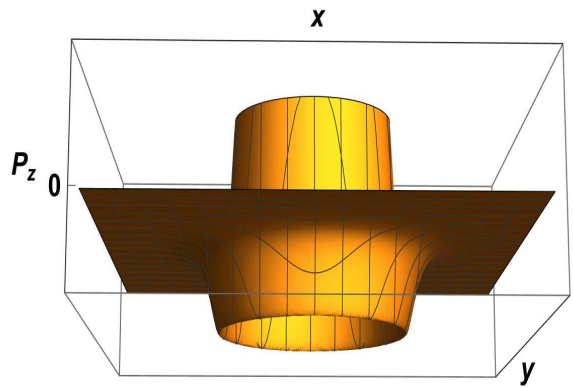}
\caption{The $z$ component of the Poynting vector for the exponential solution (\ref{exp}) of Maxwell equations. This plot is obtained for $q_z=0.1,\,\tau=0.18$ and $t=0.8$. The backflow area is clearly seen below the $P_z=0$ plane. }\label{fig4}
\end{center}
\end{figure}

\begin{figure}
\begin{center}
\includegraphics[width=0.47\textwidth,
height=0.22\textheight]{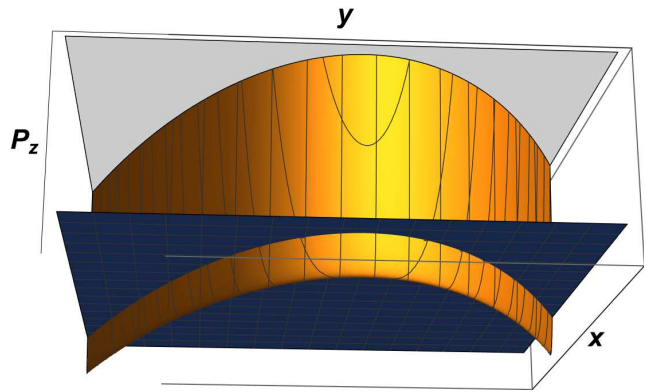}
\caption{The $z$ component of the Poynting vector for the Bessel solution of Maxwell equations obtained for $q_z=6$ and $q_\perp=0.4$. The backflow became visible only after a substantial magnification of the vertical scale.}\label{fig5}
\end{center}
\end{figure}

\begin{figure}
\begin{center}
\includegraphics[width=0.45\textwidth,
height=0.2\textheight]{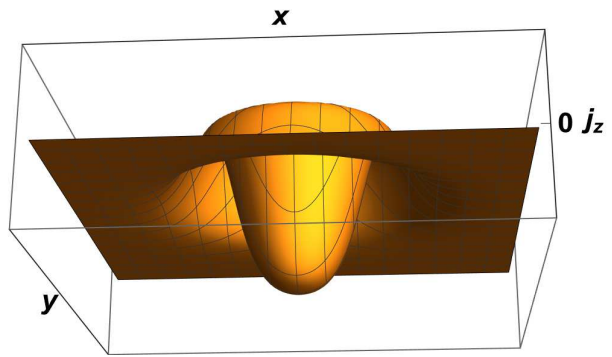}
\caption{The $z$ component of the current for the exponential solution of the Dirac equation.}\label{fig6}
\end{center}
\end{figure}
The solutions of Maxwell equations can be obtained by differentiation of a solution of the scalar wave equation \cite{rs}. The appropriate combination of derivatives is dictated by the form of the polarization vector. We choose the polarization vector again as in (\ref{ee}). The simplest solution of Maxwell equations in the family of exponential beams introduced in \cite{bb2}, expressed in terms of the RS vector is
\begin{widetext}
\begin{align}\label{exp}
\left(\begin{array}{c}F_x\\
F_y \\
F_z
\end{array}\right)
=\left(\begin{array}{c}
(\partial_t-\partial_z)^2-(\partial_x+i\partial_y)^2\\
i(\partial_t-\partial_z)^2+(\partial_x+i\partial_y)^2\\
2(\partial_t-\partial_z)(\partial_x+i\partial_y)
\end{array}\right)\
\frac{e^{iq_z z}e^{-\vert q_z\vert s}}{s},
\end{align}
\end{widetext}
where $s=\sqrt{(x^2+y^2 - (t-i\tau)^2}$, $q_z$ is the wave vector in the $z$ direction, and $\tau$ is the parameter which controls the extension of the beam in the transverse direction. The expression for the $z$ component of the Poynting vector is too long to be exhibited. We only show in Fig.~4 the final result.

Bessel beam solutions of Maxwell equations are obtained from the formula (\ref{exp}) by replacing the exponential solution by the following Bessel solution of the scalar wave equation,
\begin{align}
e^{iq_zz-i\sqrt{q_\perp^2+q_z^2}\,t}\frac{x+iy}{\sqrt{x^2+y^2}}
J_1(q_\perp\sqrt{x^2+y^2}).
\end{align}
The RS vector can be again generated according to Eq.~(\ref{exp}). The backflow is still present for this beam (see Fig.~5), but it is much less pronounced in comparison with the exponential beam.

Exponential beam solutions of the Dirac equation described by Eqs.~(15-17) of \cite{bb0} also show a prominent backflow, as is seen in Fig.~6. Bessel beam solutions described by Eq.~(11) of the same work again exhibit backflow on a very small scale.

\section{Conclusions}

We have shown that the phenomenon of backflow is a common feature of various physical system, quantum and classical, described by the linear wave equations. It follows from the linearity of the wave equations that the phenomenon of backflow does not depend on the overall amplitude of wave functions. The Fourier decomposition of these solutions into plane waves enables one to identify the direction of propagation of these elementary components. Due to the interference of these waves, in the propagation of the wave in position space we may find regions where the direction of the flow is reversed. The propagation of waves in material media (water waves, acoustic waves) will show the same phenomena as long as the linear approximation is applicable. Generalization of the backflow concept to nonlinear waves does not seem to be feasible.

\section*{Acknowledgments}

We would like to thank Radosław Łapkiewicz who aroused our interest in the phenomenon of backflow.

\appendix

\section{Backflow in the hopfion solution\\of Maxwell equations}

We have chosen here our favorite solution of Maxwell equations: the hopfion. This solution was discovered by Synge \cite{syn} and later Ra{\~n}ada \cite{ran} found its remarkable topological properties closely related to the famous Hopf fibration. Hopfion solution of Maxwell equations $\bm F_{\rm H}(\bm r,t)$ can be generated from the solution $h(\bm r,t)$ of the scalar wave equation \cite{rs,str} by differentiation,
\begin{align}\label{hopf0}
h(\bm r,t)&=\frac{1}{4\pi}
\int\!\frac{d^3k}{k}e^{-c(a+it)k}e^{i{\bm k}\cdot{\bm r}}=\frac{1}{{\bm r}^2-c^2\tau^2},\nonumber\\
\Box h(\bm r,t)&=0,
\end{align}
where $a$ is the scale parameter which determines the extension of the wave-packet in space and $\tau=t-ia$.
\begin{align}\label{ffh}
\bm F_{\rm H}=\left[\begin{array}{c}
(\partial_x+i\partial_y)^2-(\partial_t-\partial_z)^2\\
-i(\partial_x+i\partial_y)^2-i(\partial_t-\partial_z)^2\\
-2(\partial_x+i\partial_y)(\partial_t-\partial_z)
\end{array}\right]h(\bm r,t).
\end{align}

In order to tackle the problem of backflow, we have to split $h(\bm r,t)$ into two parts $h_\pm(\bm r,t)$, with positive and negative values of $k_z$,
\begin{align}\label{hopfpm}
h_\pm(\bm r,t)=\frac{1}{4\pi}\int\!\!\frac{d^3k}{k}\theta(\pm k_z)e^{-c(a+i t)k}e^{i{\bm k}\cdot{\bm r}}.
\end{align}
The Heaviside step function $\theta(\pm k_z)$ selects either positive or negative values of $k_z$.
The evaluation of these integrals is relegated to Appendix B. The resulting expressions are,
\begin{align}
h_\pm(\bm r,t)=\frac{1}{2Q(Q\mp i z)},
\end{align}
where $Q=\sqrt{\rho^2-c^2\tau^2}$ and $\rho^2=x^2+y^2$.

We can generate now the RS vector from the solution $h_+$ of the wave equation according to the prescription (\ref{ffh}). Upon the evaluation of all derivatives, we obtain the following RS vector,
\begin{align}\label{ff2}
\bm F=\frac{1}{Q^5(Q-i z)^3}\left[\begin{array}{c}
f_x\\
f_y\\
f_z
\end{array}\right],
\end{align}
where,
\begin{widetext}
\begin{align}\label{ff}
 f_x&=4\tau^4-\tau^3\zeta-\tau^2\left(8x x_+-z\zeta\right)
 +\tau\zeta\rho^2+x_+\left(4x_+\rho^2-z\zeta(x+2iy)\right),\nonumber\\
 f_y&=i\Big(4\tau^4-\tau^3\zeta+\tau^2\left(8i yx_++z\zeta\right)+\tau\zeta\rho^2
-x_+\left(4x_+\rho^2-z\zeta(2x+iy)\right)\Big),\nonumber\\
 f_z&=-8\tau^3x_++\tau^2x_+\zeta
 +\tau x_+\left(8\rho^2-3z\zeta\right)-x_+\zeta\rho^2,
\end{align}
\end{widetext}
where $x_+=x+iy$ and $\zeta=z+3iQ$.
The $z$ component of the Poynting vector is,
\begin{align}\label{pz}
P_z=\frac{18A|Q|^2+48B\Re(Q)+12C\Im(Q)+2D}{|Q|^{10}
(|Q|^2+z^2-2z\Im(Q))^3},
\end{align}
where
\begin{widetext}
\begin{align}\label{cf}
 A&=(a^2+t^2)^2(a^2+(t-z)^2)+(2a^4-2t^4+a^2(5t-z)z+t^2z(t+z))\rho^2
+(a^2+(t-z)(t+2z))\rho^4,\nonumber\\
 B&=a((a^2+t^2)^2(a^2+t(t-2z))+2(a^2-t^2)(a^2+t(t+z))\rho^2+(a^2+t(t+4z))\rho^4),\nonumber\\
 C&=(a^2+t^2)^2(a^2(4t+3z)+(t-z)(4t^2-tz+z^2))+(8a^4t+a^2z(20t^2-5tz+z^2)
-t^2(8t^3-4t^2z+tz^2+z^3))\rho^2\nonumber\\
&+(a^2(4t-9z)+(t+2z)(4t^2-tz+z^2))\rho^4-6z\rho^6,\nonumber\\
 D&=(a^2+t^2)^2(16a^4+a^2(32t^2-8tz-7z^2)+(4t^2-tz+z^2)^2)
+(32(a-t)(a+t)(a^2+t^2)^2+16t(t^4-a^4)z\nonumber\\
&-2(a^4+20a^2t^2+3t^4)z^2
+t(5a^2+t^2)z^3+(t^2-a^2)z^4)\rho^2+z(a^2(17z-8t)-(t+2z)(8t^2-tz+z^2))\rho^4\nonumber\\
&+4(8t^2+3z^2-8a^2)\rho^6-16\rho^8.
\end{align}
\end{widetext}
In Fig.7 the region where ${\mathcal G}_z<0$ is clearly seen.

\begin{figure}
\begin{center}
\includegraphics[width=0.47\textwidth,
height=0.3\textheight]{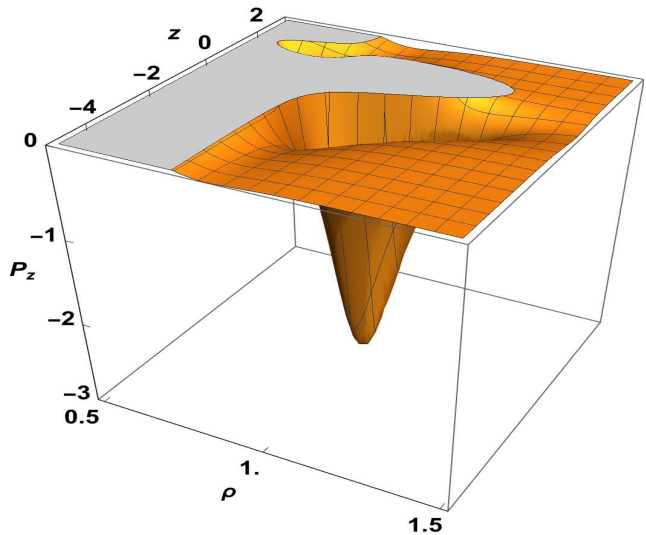}
\caption{This plot of the $z$ component of the Poynting vector for the hopfion solution is obtained for $t=0.8\;a=0.2$. Space coordinates are measured in units of $ca$. The scale for ${\mathcal G}_z$ is arbitrary because the electromagnetic field can have any strength. The light gray area marks the region in the $\rho-z$ plane where ${\mathcal G}_z$ changes sign.}\label{fig7}
\end{center}
\end{figure}

\section{Evaluation of the Fourier integrals}

The evaluation of the integrals (\ref{hopfpm}) proceeds in three steps. Since the $z$ direction is distinguished, we choose the cylindrical coordinates $\rho,\phi$ and $k_z$. The integration over $\phi$ produces the Bessel function $2\pi J_0(\kappa\rho)$, where $\kappa=\sqrt{k_x^2+k_y^2}$. The integration over $\kappa$ in the resulting integral,
\begin{align}
h_\pm(\rho,z,t)&=\frac{1}{2}\int_{-\infty}^{\infty}
\!\!dk_z\theta(\pm k_z)e^{i k_z z}\\
&\times\int_0^{\infty}\!\!\frac{d\kappa\,\kappa}{\sqrt{\kappa^2+k_z^2}}J_0(\kappa\rho) e^{-c(a+it)\sqrt{\kappa^2+k_z^2}},\nonumber
\end{align}
can be done with the help of the formula 6.646 of \cite{gr}. All that is required is the change of the integration variable $x$ in this formula and the appropriate choice of parameters $\alpha$ and $\beta$. The result is,
\begin{align}
&\int_0^{\infty}\!\!\frac{d\kappa\,\kappa}{\sqrt{\kappa^2+k_z^2}}J_0(\kappa\rho) e^{-c(a+it)\sqrt{\kappa^2+k_z^2}}\nonumber\\
&=e^{-|k_z|\sqrt{\rho^2-c^2\tau^2}}/\sqrt{\rho^2-c^2\tau^2}.
\end{align}
In the last step we do a simple integration over $k_z$ and obtain,
\begin{align}
h_\pm(\rho,z,t)=\frac{1}{2\sqrt{\rho^2-c^2\tau^2}(\sqrt{\rho^2-c^2\tau^2}\mp iz)}
\end{align}
As a final check, we have,
\begin{align}
h_+(\rho,z,t)+h_-(\rho,z,t)=h(\rho,z,t)
\end{align}

\end{document}